\documentclass[%
 reprint,
 prc,
 superscriptaddress,
 amsmath,amssymb,
]{revtex4-2}

\usepackage{siunitx}
\usepackage{multirow}
\usepackage{color}

\usepackage{graphicx}
\usepackage{dcolumn}
\usepackage{bm}
\usepackage[mathlines]{lineno}

\begin{document}


\title{Comprehensive mass measurement study of ${}^{252}$Cf fission fragments with MRTOF-MS \\ and  detailed study of masses of neutron-rich Ce isotopes}

\author{S.~Kimura}
\email[]{sota.kimura@kek.jp}
\affiliation{Wako Nuclear Science Center (WNSC), Institute of Particle and Nuclear Studies (IPNS), High Energy Accelerator Research Organization (KEK), Wako 351-0198, Japan}
\affiliation{RIKEN Nishina Center for Accelerator-Based Science, Wako 351-0198, Japan}

\author{M.~Wada}
\affiliation{Wako Nuclear Science Center (WNSC), Institute of Particle and Nuclear Studies (IPNS), High Energy Accelerator Research Organization (KEK), Wako 351-0198, Japan}

\author{H.~Haba}
\affiliation{RIKEN Nishina Center for Accelerator-Based Science, Wako 351-0198, Japan}

\author{H.~Ishiyama}
\affiliation{RIKEN Nishina Center for Accelerator-Based Science, Wako 351-0198, Japan}

\author{S.~Ishizawa}
\affiliation{Institute for Materials Research (KINKEN, IMR), Tohoku University, Sendai 980-8577, Japan}
\affiliation{New Industry Creation Hatchery Center (NICHe), Tohoku University, Sendai 980-8579, Japan.}

\author{Y.~Ito}
\affiliation{Advanced Science Research Center, Japan Atomic Energy Agency (JAEA), Tokai, 319-1195, Japan.}
\affiliation{RIKEN Nishina Center for Accelerator-Based Science, Wako 351-0198, Japan}

\author{T.~Niwase}
\affiliation{Department of Physics, Kyushu University, Fukuoka, 819-0395, Japan}

\author{M.~Rosenbusch}
\affiliation{RIKEN Nishina Center for Accelerator-Based Science, Wako 351-0198, Japan}

\author{P.~Schury}
\affiliation{Wako Nuclear Science Center (WNSC), Institute of Particle and Nuclear Studies (IPNS), High Energy Accelerator Research Organization (KEK), Wako 351-0198, Japan}

\author{A.~Takamine}
\affiliation{RIKEN Nishina Center for Accelerator-Based Science, Wako 351-0198, Japan}

\date{\today}

\begin{abstract}
We report the mass measurements of neutron-rich isotopes produced via spontaneous fission of ${}^{252}{\rm Cf}$ using a multi-reflection time-of-flight mass spectrograph. The mass of ${}^{155}{\rm Ce}$ is determined experimentally for the first time. A discrepancy between the experimental and literature values was found for the mass of ${}^{127}{\rm Sb}$, which was previously deduced through indirect measurements. In comparison with several theoretical predictions, both the values and the trend of the mass excesses of ${}^{152-155}{\rm Ce}$ cannot be consistently explained. The wide-range and simultaneous mass measurements of the multi-reflection time-of-flight mass spectrograph enable us to cross-check the existing mass data, and the conflict between the measured time-of-flight ratio and the extracted mass would imply the necessity of reexamining them.  
\end{abstract}

\maketitle



\section{Introduction}

\begin{figure}[t]
\includegraphics[width=0.5\textwidth, bb=0 0 842 595, clip, trim= 25 30 30 25]{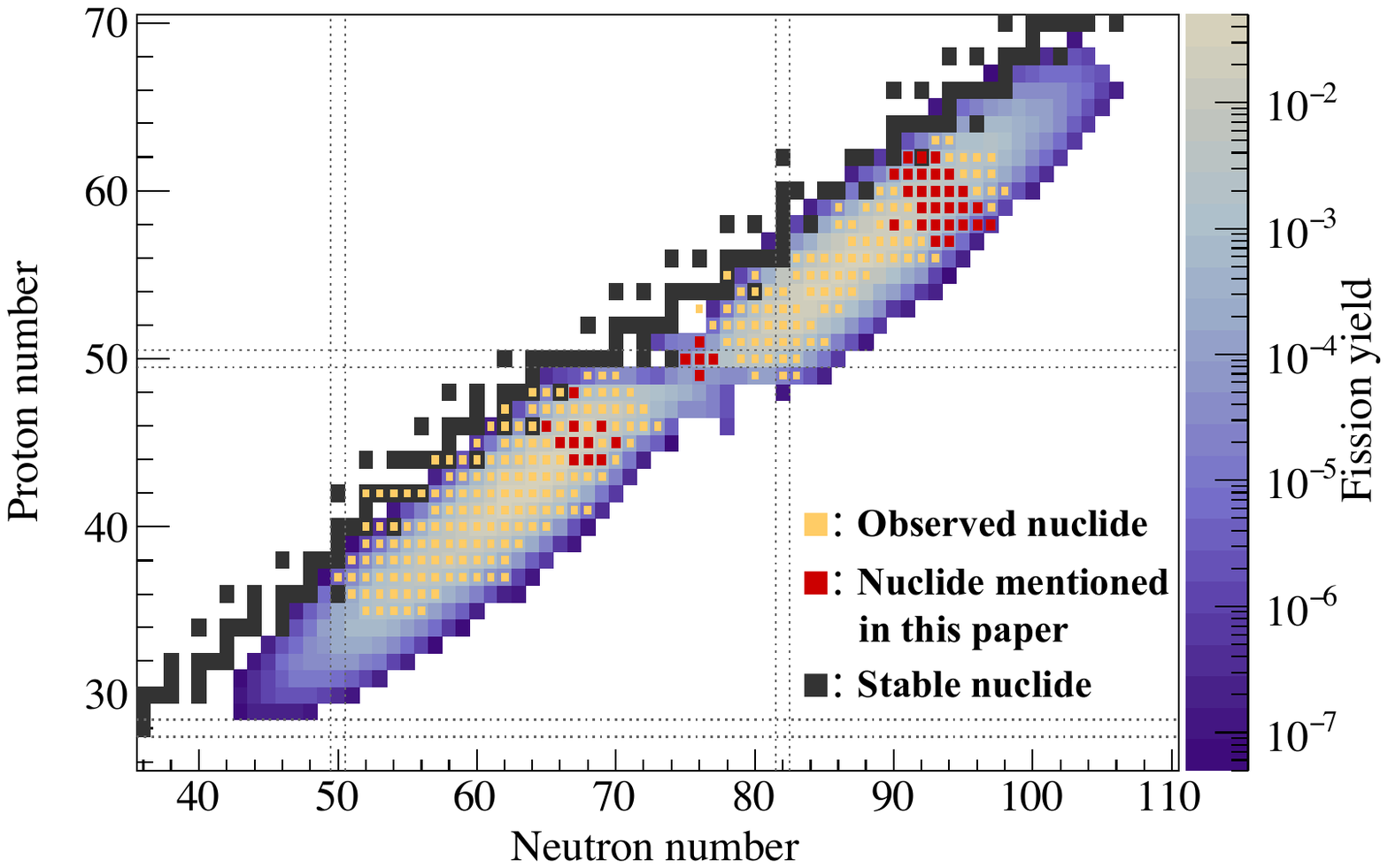}
\caption{\label{NuclChart}Location of the observed nuclides on the nuclear chart. The observed nuclides are indicated with orange-colored squares. The black squares represent the stable nuclides. The nuclides mentioned in this paper are marked with red-colored squares. The contour map shows the fission yield of $^{252}$Cf \citep{Katakura2012}.}
\end{figure}

Nuclear mass measurements provide essential data for nuclear physics to reveal the underlying nuclear structure and to understand the nuclear reactions \citep{Lunney2003, Dilling2018}. Particularly for nuclear astrophysics, a large set of accurate and precise mass data is required to understand the origin of elements. The rapid neutron-capture ($r$-) process path is governed by the balance between the $\beta^{-}$-decay rate and neutron-capture rate. In the neutron-capture process, $({\rm n},\gamma)-(\gamma,{\rm n})$ equilibrium will be established, and the net neutron-capture rate strongly depends on the $Q$-value. Nuclear masses are crucial in determining the final isotopic abundance of the $r$-process. Access to the $r$-process nuclei is difficult because their locations are far from stability; thus, theoretical predictions are needed to obtain the final isotopic abundances. A mass precision better than $\sim$50~keV is required to determine the isotopic abundance ratio to be better than a factor of 2 \cite{Schatz2013}. However, in the regions without experimental data on nuclear masses, there is a wide variance amongst the masses predicted by various theoretical models \citep{Mumpower2015}. 

In the region of neutron-rich lanthanoid isotopes, many nuclear masses have been determined with Penning trap mass spectrometers (PT-MSs). Amongst these measurements of the neutron-rich isotopes, which can be produced by the spontaneous fission of ${}^{252}{\rm Cf}$, there is a significant contribution from the Canadian Penning trap (CPT) at the CARIBU facility \citep{Savard2006, Schelt2012, Orford2018, Orford2020, Orford2022}. Mass measurement studies focusing on the same region with JYFLTRAP at Jyv\"askyl\"a have also been reported \citep{Vilen2018, Vilen2020E, Vilen2020}. The seven nuclides and one isomeric state, $^{154,156,158}$Nd, $^{162,163}$Sm, $^{162}$Eu$^{g,m}$, and $^{163}$Gd are reported in the studies with both JYFLTRAP and CPT. Based on the updated result of JYFLTRAP \citep{Vilen2020E, Vilen2020}, for the nuclides excluding $^{156,158}$Nd, the extracted mass excesses of both studies are in agreement. Both facilities measured the excitation energy of $^{162}$Eu$^{m}$, and their results are slightly different, $\Delta E_{\rm X, CPT-JYFL} ({}^{162}{\rm Eu}{}^{m})= 4.2(37)~{\rm keV}$, while the ground state's mass excess values are consistent. For the remaining two Nd isotopes, one can find that the differences in their mass excess values are $\Delta {\rm ME}_{\rm CPT-JYFL}({}^{156}{\rm Nd}) = 7.9(24)~{\rm keV}$ and $\Delta {\rm ME}_{\rm CPT-JYFL}({}^{158}{\rm Nd}) = 62(37)~{\rm keV}$. While the difference for $^{156}$Nd is less than 10 keV, the $^{158}$Nd case shows more than 50~keV, but its deviation remains at a level of $1.7\sigma$ due to the relatively large uncertainty provided by JYFLTRAP.  This is significant considering the requirement of the $r$-process calculation. For most isotopes in the neutron-rich lanthanoid region, the presently accepted masses, as determined by the 2020 Atomic Mass Evaluation (AME20) \citep{Huang2021, Wang2021}, are based on single mass measurement. Thus, cross-checking the existing results with the measurements via other techniques is desired to improve the accuracy of the nuclear mass data.  


Figure \ref{NuclChart} shows the observed isotopes in our study of $^{252}$Cf fission fragments with a multi-reflection time-of-flight mass spectrograph (MRTOF-MS). More than 250 isotopes covering 29 elements have been extracted from the gas cell and observed in the time-of-flight (TOF) spectra. We report the results of these measurements, a highlight of which are neutron-rich lanthanoid isotopes extending to $^{155}$Ce.

\section{Experiment}

\begin{figure}[t]
\includegraphics[width=0.45\textwidth, bb=0 0 842 595, clip, trim=100 60 90 70]{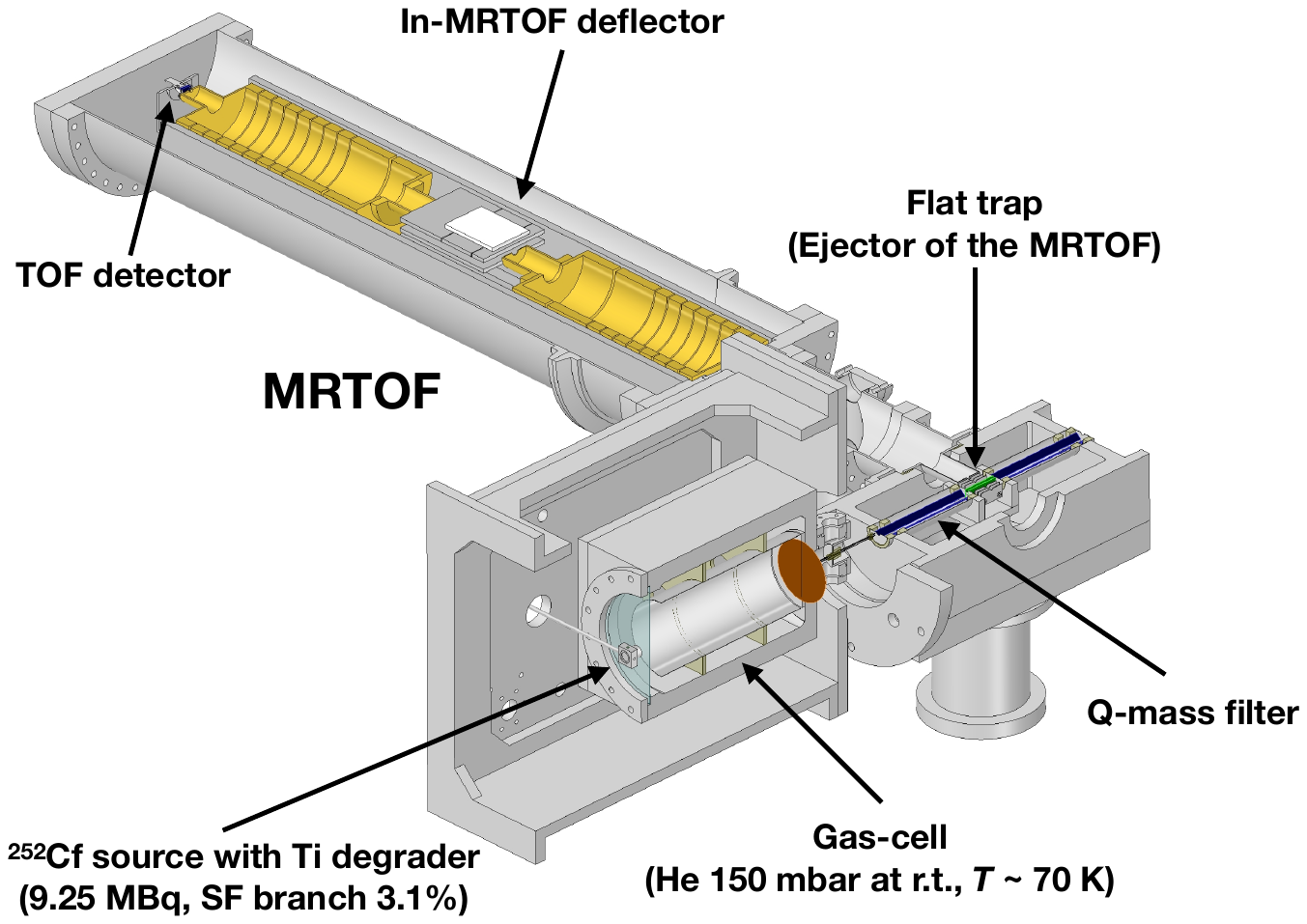}
\caption{\label{Setup}Schematic view of the experimental setup.}
\end{figure}

Figure \ref{Setup} shows a schematic of the experimental setup of the present study. The experimental setup comprises a cryogenic He gas cell (GC), an ion transport suite, and the MRTOF-MS. A 9.25~MBq $^{252}$Cf source (spontaneous fission branch: 3.1\% \citep{Kondev2021}) with a 6-$\mu$m Ti degrader is installed in front of the GC. The GC has a 1.5-$\mu$m Mylar window and is cryogenically cooled to $\sim$$70~{\rm K}$ throughout all measurements. The He gas pressure is regulated to maintain a density equivalent to 150 mbar at room temperature. The emitted fission fragments from the $^{252}$Cf source will be stopped and thermalized inside the GC. The thermalized ions are transported by an RF-carpet \citep{Arai2014} toward the exit of the GC and further to a planar-geometry radio-frequency quadrupole (RFQ) trap (flat trap) \citep{Ito2013b}, which works as an ejection device to the MRTOF-MS, via the ion transport system having two linear RFQ traps. The second linear RFQ trap can be used as a quadrupole mass filter and enable the removal of the intense contaminants produced inside the GC via ionization with the $\alpha$-particles emitted by the $^{252}$Cf source.

The transported analyte ions are perpendicularly ejected from the flat trap to the MRTOF-MS. The ejection trigger signals are also used as the start signal of a time-to-digital converter (TDC) start trigger. The ejected ions are captured by temporarily lowering the entrance mirror potential of the MRTOF-MS and raising it before the ions return to the entrance mirror. After the ions make a certain number of reflections, making the time focus condition, their TOFs are measured by lowering the exit mirror potential and allowing the ions to fly to the TOF detector. 

The flat trap has a wide mass acceptance, making it capable of accumulating and storing across a wide range of $A/q$ values simultaneously, while the MRTOF-MS similarly has a broad acceptance to capture ions spanning a wide $A/q$ range. This can lead to a contaminated TOF spectrum with many unwanted ion species. To avoid this, an in-MRTOF deflector (IMD) \citep{Rosenbusch2023} has been installed at the middle point, the drift part of the MRTOF-MS. The IMD  consists of a pair of two planar electrodes and can remove unwanted ions by applying a voltage of 20~V when the analyte ions are not passing through the IMD. The operation condition of the IMD  is determined with $^{85}{\rm Rb}^{+}$ ions from an offline ion source to calibrate the timing of the IMD  pulses and ensure that desired ion species are not disturbed. 

\section{Analysis}

The analysis strategy is similar to the previous work with the MRTOF-MS \citep{Kimura2018}. The measured TOF value for an ion, having mass $m$ and charge $q$,  which undergoes $n$ laps in the MRTOF-MS device, can be represented by 
\begin{equation}
t_{\rm obs} = (a + b \cdot n) \sqrt{m/q} + t_0
\label{eqObsTOF}
\end{equation}
where $a$ and $b$ are constants related to the non-reflection flight path and the path between consecutive reflections, respectively, and $t_0$ represents an electronic delay between the TDC start signal and the ion's actual ejection from the flat trap. The single reference method \cite{Ito2013}, which needs only one reference mass, is adopted to determine the mass of analyte ions. In this method $m_{\rm X}$, the ionic mass of nuclide X, is given by Eq.~\ref{eqSingleRef}:
\begin{equation}
m_{\rm X} = \rho^2 m_{\rm ref} = \left( \frac{t_{\rm X}-t_0}{t_{\rm ref}-t_0} \right)^2 m_{\rm ref},
\label{eqSingleRef}
\end{equation}
where $\rho$ is the TOF ratio, $t_{\rm X}$ and $t_{\rm ref}$ are the TOF of nuclide X and the reference ion, respectively, $m_{\rm ref}$ is the mass of the reference ion,  $t_0$ is the constant time offset within the measurement system mentioned above. As the mass reference, we select the nuclide, which has no known long-lived isomeric states and shows the highest intensity spectral peak.

Collecting sufficient numbers of the more exotic nuclides in an isobar chain, the required measurement time was over twelve hours for each. During this time, the TOF could vary considerably due to the thermal expansion of the MRTOF-MS device and minor instabilities in the high-voltage power supply system for the mirror electrodes \citep{Schury2014}. In all measurements of the present study, the fission fragment and the ${}^{85}{\rm Rb}$ ions provided from an offline ion source were alternately measured in 25-ms cycles. Then, these TOF drifts can be compensated by the use of ${}^{85}{\rm Rb}$ ions. The TOF corrections were performed within each subset, obtained by dividing the raw data set into ${\cal N}$ parts. For ions in each subset $i$ the corrected TOF $t_{\rm corr, i}$ is  calculated using  the following relation:
\begin{equation}
t_{\rm corr, i} = t_{\rm raw,i} \left( \frac{t_{\rm 85Rb}}{t_{\rm 85Rb,i}}\right),
\end{equation}
where $t_{\rm raw,i}$ is the uncorrected TOF of each ion in subset $i$, while $ t_{\rm 85Rb}$ and $t_{\rm 85Rb,i}$ are the standard TOF of ${}^{85}{\rm Rb}$ and the fitted TOF center of the $i^{th}$ spectral subset, respectively. This correction method was applied to all the data presented here. 

To fully preclude accidental misattribution in identifying the ions in the TOF spectra, the measurements were performed at two different numbers of laps, 602 and 604, for each $A/q$ series. The final results of two measurements belonging to the same $A/q$ series were obtained as their weighted average,
\begin{eqnarray}
\overline{\rho^2} &=& \frac{w^2_{602} \rho^2_{602}+w^2_{604} \rho^2_{604}}{w^2_{602}+w^2_{604}}, \\
\delta (\overline{\rho^2}) &=& \frac{1}{\sqrt{w^2_{602}+w^2_{604}}},
\end{eqnarray}
where $w_i$ is a weight of each measurement and is defined by $w_i \equiv 1/ \delta (\rho_i^2)$.

The atomic mass of nuclide X, $M_{\rm X}$, observed as the $q_{\rm X}+$ ions, is given by 
\begin{equation}
M_{\rm X} = q_{\rm X}(m_{\rm ref}\overline{\rho^2}  + m_{\rm e}),
\end{equation}
where $m_{\rm ref}$ and $m_{\rm e}$ are the ionic masses of the references and the electron rest mass, respectively. \\

A phenomenological fitting function, based on an exponential-Gaussian hybrid function \citep{Ito2013, Schury2014, Lan2001}, was used to fit non-Gaussian-shape peaks accurately. In the present study, we employed the function:
\begin{align}
f(\tau) = \left\{
\begin{array}{l}
\left( \kappa / \sigma \right) \exp \left[ \frac{t_{\rm s1} (t_{\rm s1} -2\tau)}{2\sigma^2} \right] 
\ \left({\rm for} \ \tau < t_{\rm s1}\right), \\
\\
\left( \kappa / \sigma \right) \exp \left[ - \frac{\tau^2}{2\sigma^2} \right] 
\ \left({\rm for} \ t_{\rm s1} \leq \tau < t_{\rm s2}\right)\\
\\
\left( \kappa / \sigma \right) \exp \left[ \frac{t_{\rm s2} (t_{\rm s2} -2\tau)}{2\sigma^2} \right] 
\ \left({\rm for} \ \tau \geq t_{\rm s2}\right),
\end{array}
\right.
\end{align}
where $t_{{\rm s}i}$ denotes  the range of each sub-function. The variable $\tau$ is defined as $\tau \equiv t - \mu$, where $\mu$ is the peak center used in the mass determinations. 

Herein, we set an assumption about the peak shape: the peaks of the ions belonging to the same $A/q$ series have an identical shape. Based on this assumption, the only free parameters for each peak in the fitting function are the peak center $\mu$ and the peak height $\kappa$ for the species of interest. In the fitting algorithm, to improve the mass precision, the $\tau$ parameter was treated as a function of $t$, $t_{\rm ref}$, and $\rho$, where $\rho$ is the TOF ratio from Eq.~\ref{eqSingleRef}. Then, the fitting function $F$ for $N$ peaks was described by 
 \begin{eqnarray}
F(t, t_{\rm ref}, \rho_1, \cdots, \rho_{\rm N},, \kappa_1, \cdots, \kappa_{\rm N}) =  \nonumber  \\
\sum_{i=1}^{N} f(t, t_{\rm ref}, \rho_i,\kappa_i, \sigma, t_{{\rm s}1}, t_{{\rm s}2}).
\end{eqnarray}
The ROOT package \citep{Brun1997} was used for peak fitting. There is an ambiguity in the degree of freedom in the binning process to fit: the width and offset of the center. To take their influence into account, the systematic error of $\delta \rho_{\rm sys, bin} = 10^{-8}$ \citep{Kimura2018} was added to the error of each measured $\rho$ as an additional error.

\begin{figure}[t]
\includegraphics[width=0.45\textwidth, bb=0 0 842 595, clip, trim=205 0 205 0]{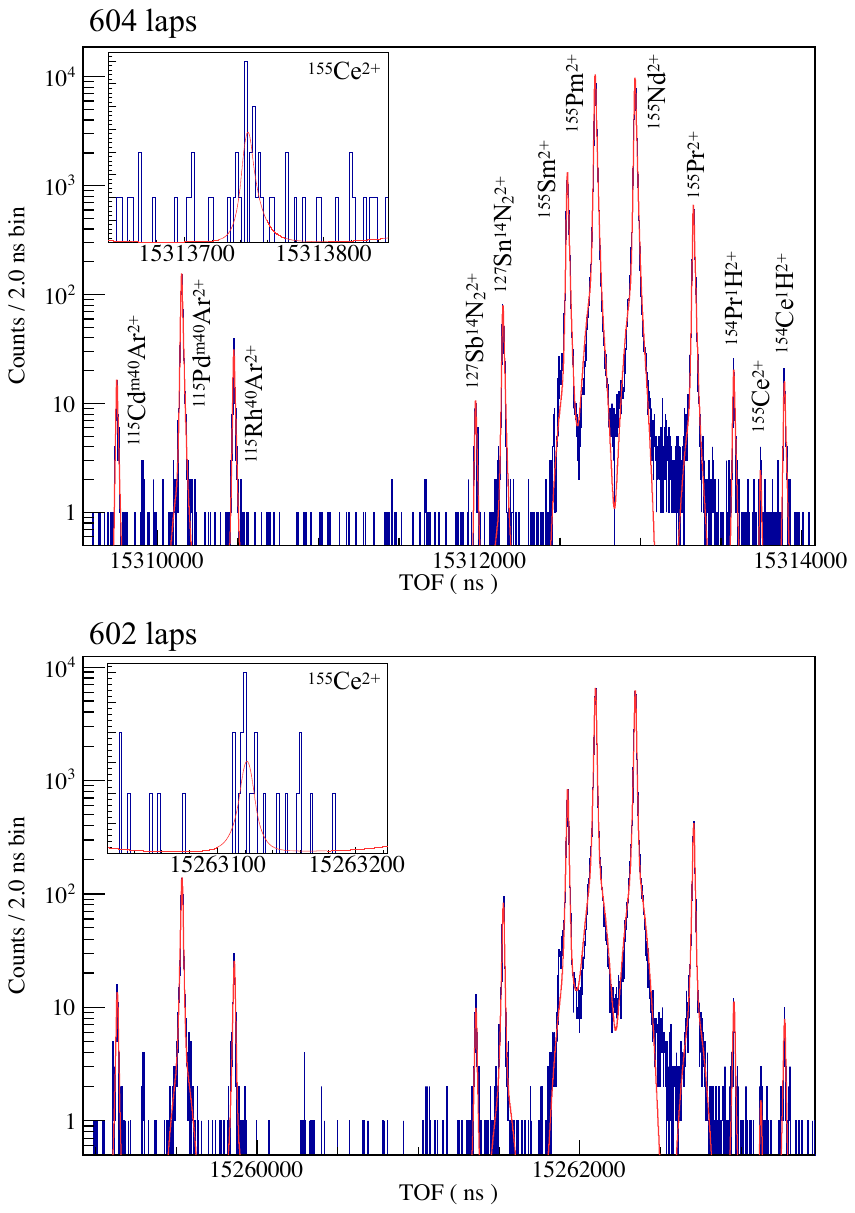}
\caption{\label{A=155++}TOF spectra of $A/q=77.5$ series. Red lines show the fit results with the phenomenological fitting function. The enlarged spectra around the predicted positions of $^{155}$Ce are indicated in the top-left panels.}
\end{figure}

\begin{figure*}[p]
\setcounter{figure}{3}
\includegraphics[width=0.9\textwidth, bb=0 0 595 842, clip, trim=0 0 0 0]{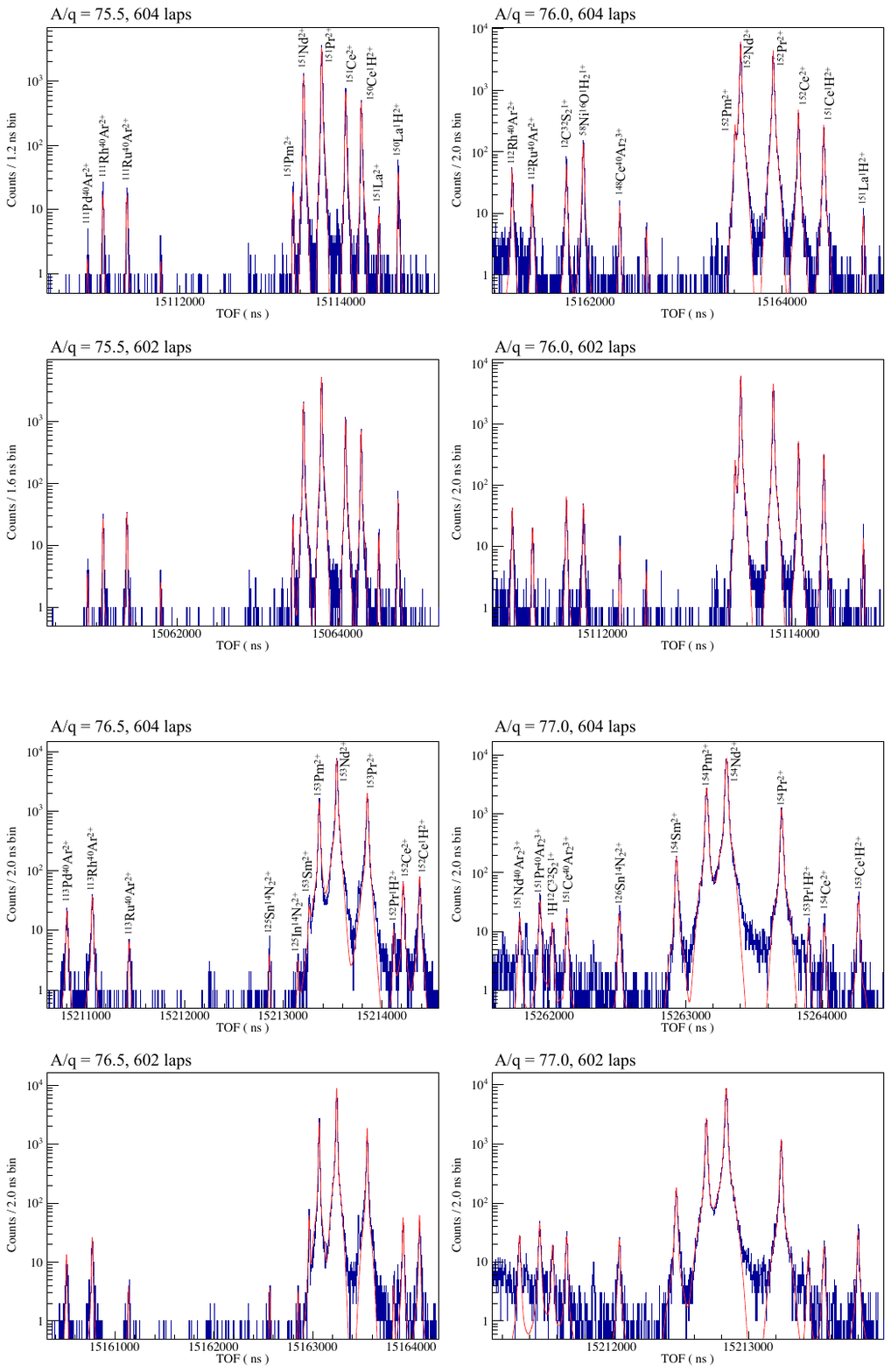}
\caption{\label{A=151-154}TOF spectra of $A/q=75.5 - 77.0$ series. Red lines show the fit results with the phenomenological fitting function.}
\end{figure*}

\section{Results and Discussion}

In the present paper, we focus on the five $A/q$ series: $A/q = 75.5$, 76.0, 76.5, 77.0, and 77.5. The TOF spectra of $A/q=77.5$ series at 602 and 604 laps are shown in Fig.~\ref{A=155++}. Other $A/q$ series' TOF spectra are given in Fig.~\ref{A=151-154}. If the relative positions and intensities of a given set of peaks remain constant at different numbers of laps, we can infer that they are isobaric with each other. In the TOF spectra, we observed not only atomic but also molecular adduct ions with the same/different charge states, $e.g.$, $^{14}{\rm N}_2$-attached $2+$ ions and $^{40}{\rm Ar}_2$-attached $3+$ ions, and stable molecular ions. Thus, many ions with different mass numbers $A$ can be included in the same TOF spectrum. This allows us to confirm a mass correlation between not only two ion species but also all mass correlations with all other observed species simultaneously. This is an excellent advantage of the MRTOF-MS for achieving accurate measurements. 
 
In the present measurements, the mass resolving power was $R_{\rm m} \sim 500,000$. Some observed isotopes have low-lying isomeric states with an excitation energy of $E_{\rm X} \lesssim 100~{\rm keV}$. They cannot be resolved based on the present mass resolving power; we treated them as one peak with a single component. 

The accuracy of mass determination by the MRTOF-MS is at the $3.5 \times 10^{-8}$ level when an accurate isobaric reference is used, as in the previous study \citep{Kimura2018}. The systematic mass error resulting from the uncertainty of the constant time offset $\delta t_0$ is estimated to be $\delta(\rho^2)/\rho^2 \lesssim 10^{-9}$ when employing isobaric references \citep{Itothesis, Kimura2021} and is negligible in the present study. The most intense peak of the fission products, i.e., a nuclide previously measured with PT-MS, was used as the reference for each $A/q$. Specifically, ${}^{151}$Pr ($\delta m_{\rm ref}  = 12$~keV), ${}^{152}$Pr (9.8~keV), ${}^{153}$Nd (2.7~keV), ${}^{154}$Nd (1.0~keV), and ${}^{155}$Pm (4.7~keV) were used as the mass references. We found that if the uncertainty of a reference is smaller than a few keV, it can cause as much as a $3\sigma$ discrepancy for the mass value of another previously measured isobar. The MRTOF-MS's high linearity of TOF regarding ion mass suggests a problem with the error evaluation in, at least, one of the existing mass data. However, at this point, it is unclear which one is correct. Thus, to extract reasonable mass excess while maintaining the consistency of the measured TOF ratio, which is equivalent to the ion mass ratio, we modified the uncertainties of the mass references. They are specifically ${}^{153}$Nd ($\delta m_{\rm ref} = 2.7$~keV$\rightarrow$16~keV), ${}^{154}$Nd (1.0~keV$\rightarrow$13~keV), and ${}^{155}$Pr (4.7~keV$\rightarrow$7.1~keV).

The results are summarized in Table~\ref{summary1} and \ref{summary2}. In the following parts, we will discuss the details of the results of each $A/q$ series. 
\begin{table*}[p]
\setcounter{table}{0}
\caption{ \label{summary1}Measured values of the squares of time-of-flight ratio $\rho^2$ and the mass excess values ME. The nuclides used as atomic mass references are shown in column ``Ref.". ME$_{\rm lit}$ indicates the literature's mass excess values, and $\Delta {\rm ME}$ represents the differences between ME$_{\rm lit}$ and the mass excess values of the present study: $\Delta {\rm ME} \equiv {\rm ME} - {\rm ME}_{\rm lit}$. The reference of ME$_{\rm lit}$ is indicated in column ``Ref. of lit."; AME20 \citep{Huang2021, Wang2021} and CPT \citep{Orford2022}. The $\delta$m/m column provides relative mass precisions of the present measurements. The mass reference's modified uncertainty is presented with an arrow connected to the original value.}
\renewcommand{\arraystretch}{1.5}
\begin{ruledtabular}
\begin{tabular}{cllSScS} 
\textrm{Species}&
\multicolumn{1}{c}{\textrm{$\rho^2$}}&
\multicolumn{1}{c}{\textrm{$\delta (\rho^2) / \rho^2$}}&
\multicolumn{1}{c}{\textrm{ME~(keV)}}&
\multicolumn{1}{c}{\textrm{ME$_{\rm lit.}$~(keV)}}&
\multicolumn{1}{c}{\textrm{Ref.~of~lit.}}&
\multicolumn{1}{c}{\textrm{$\Delta {\rm ME}$~(keV)}} 
\\ \hline
\colrule
%
%
$A/q = 75.5$&&&&&\\
${}^{111}{\rm Pd}$${}^{40}{\rm Ar}^{2+}\footnote[1]{Admixture of the ground state and the isomeric state(s)}$&0.99961500(13)&$1.3 \times 10^{-7}$&-120906(30)&-121025.79(73)&AME20&120(30)\\
${}^{111}{\rm Rh}$${}^{40}{\rm Ar}^{2+}$&0.999640390(47)&$4.7 \times 10^{-8}$&-117336(24)&-117343.7(6.9)&AME20&7(25)\\
${}^{111}{\rm Ru}$${}^{40}{\rm Ar}^{2+}$&0.999679667(46)&$4.6 \times 10^{-8}$&-111815(24)&-111825.2(9.7)&AME20&11(26)\\

${}^{151}{\rm Pm}^{2+}$&0.999953012(47)&$4.7 \times 10^{-8}$&-73386(24)&-73386.3(4.6)&AME20&1(25)\\
${}^{151}{\rm Nd}^{2+}$&0.999970367(18)&$1.8 \times 10^{-8}$&-70946(23)&-70943.2(1.1)&AME20&-3(24)\\
${}^{151}{\rm Pr}^{2+}$&1&&&-66780(12)&AME20&\\
${}^{151}{\rm Ce}^{2+}\footnotemark[1]$&1.000039526(19)&$1.9 \times 10^{-8}$&-61223(24)&-61225(18)&AME20&2(29)\\
${}^{150}{\rm Ce}$${}^{1}{\rm H}^{2+}$&1.000065422(20)&$2.0 \times 10^{-8}$&-57582(24)&-57558(12)&AME20&-24(26)\\
${}^{151}{\rm La}^{2+}$&1.000094199(89)&$8.9 \times 10^{-8}$&-53537(26)&-53310(435)&AME20&-226(436)\\
${}^{150}{\rm La}$${}^{1}{\rm H}^{2+}$&1.000126206(35)&$3.5 \times 10^{-8}$&-49037(24)&-49022.1(2.5)&AME20&-15(24)\\
%
%
\hline
$A/q = 76.0$&&&&&\\
${}^{112}{\rm Rh}$${}^{40}{\rm Ar}^{2+}$\footnotemark[1]&0.999640920(55)&$5.5 \times 10^{-8}$&-114600(21)&-114771(44)&AME20&171(49)\\
${}^{112}{\rm Ru}$${}^{40}{\rm Ar}^{2+}$&0.999668761(51)&$5.1 \times 10^{-8}$&-110660(21)&-110670.7(96)&AME20&11(23)\\
${}^{12}{\rm C}{}^{32}{\rm S}_2^{1+}$&0.999715477(32)&$3.3 \times 10^{-8}$&-52024(10)&-52031.0743(26)&AME20&7(10)\\
${}^{58}{\rm Ni}{}^{16}{\rm O}{}^{1}{\rm H}_2^{1+}$&0.999738692(29)&$2.9 \times 10^{-8}$&-50382(10)&-50387.93(35)&AME20&6(10)\\
${}^{148}{\rm Ce}$${}^{40}{\rm Ar}_2^{3+}$&0.99978904(12)&$1.2 \times 10^{-7}$&-140456(39)&-140478(11)&AME20&22(41)\\

${}^{152}{\rm Pm}^{2+}$\footnotemark[1]&0.999947464(25)&$2.5 \times 10^{-8}$&-71217(20)&-71254(26)&AME20&37(33)\\
${}^{152}{\rm Nd}^{2+}$&0.999955016(17)&$1.7 \times 10^{-8}$&-70148(20)&-70150(25)&AME20&1(31)\\
${}^{152}{\rm Pr}^{2+}$&1&&&-63782.2(98)&CPT&\\
${}^{152}{\rm Ce}^{2+}$&1.000034560(18)&$1.8 \times 10^{-8}$&-58891(20)&-58878.3(23)&CPT&-13(20)\\
${}^{151}{\rm Ce}$${}^{1}{\rm H}^{2+}$\footnotemark[1]&1.000069506(20)&$2.0 \times 10^{-8}$&-53945(20)&-53936(18)&AME20&-9(27)\\
${}^{151}{\rm La}$${}^{1}{\rm H}^{2+}$&1.000123838(62)&$6.2 \times 10^{-8}$&-46256(22)&-46021(435)&AME20&-235(436)\\

%
%
 \hline
$A/q = 76.5$&&&&&\\
${}^{113}{\rm Pd}$${}^{40}{\rm Ar}^{2+}$&0.99963982(15)&$1.5 \times 10^{-7}$&-118639(39)&-118630.4(70)&AME20&-8(40)\\
${}^{113}{\rm Rh}$${}^{40}{\rm Ar}^{2+}$&0.999673917(48)&$4.8 \times 10^{-8}$&-113781(34)&-113806.8(71)&AME20&26(34)\\
${}^{113}{\rm Ru}$${}^{40}{\rm Ar}^{2+}$\footnotemark[1]&0.99972269(20)&$2.0 \times 10^{-7}$&-106833(44)&-106908(38)&AME20&75(58)\\

${}^{125}{\rm Sn}^{\rm m}$${}^{14}{\rm N}_2^{2+}$\footnotemark[1]&0.99991043(16)&$1.6 \times 10^{-7}$&-80089(40)&-80139.3(13)&AME20&23(40)\\
${}^{125}{\rm In}$${}^{14}{\rm N}_2^{2+}$&0.99994834(18)&$1.8 \times 10^{-7}$&-74690(42)&-74685.5(1.8)&AME20&-5(42)\\

${}^{153}{\rm Sm}^{2+}$&0.999963232(36)&$3.6 \times 10^{-8}$&-72568(33)&-72560.1(10)&AME20&-8(33)\\
${}^{153}{\rm Pm}^{2+}$&0.999976473(18)&$1.8 \times 10^{-8}$&-70682(33)&-70648.0(91)&AME20&-34(34)\\
${}^{153}{\rm Nd}^{2+}$&1&&&\multicolumn{1}{c}{\hspace{1.8em}$-67\hspace{0.2em}330.4(27 \rightarrow 160)$}&AME20&\\
${}^{153}{\rm Pr}^{2+}$&1.000040537(15)&$1.5 \times 10^{-8}$&-61556(33)&-61547.5(24)&AME20&-8(33)\\
${}^{152}{\rm Pr}$${}^{1}{\rm H}^{2+}$&1.00007585(15)&$1.5 \times 10^{-7}$&-56526(39)&-56493.2(98)&CPT&-33(40)\\
${}^{153}{\rm Ce}^{2+}$&1.000088541(34)&$3.4 \times 10^{-8}$&-54718(33)&-54711.5(24)&CPT&-6(33)\\
${}^{152}{\rm Ce}$${}^{1}{\rm H}^{2+}$&1.000110410(36)&$3.6 \times 10^{-8}$&-51602(33)&-51589.3(2.3)&CPT&-13(33)\\

\end{tabular}
\end{ruledtabular}
\renewcommand{\arraystretch}{1.0}
\end{table*}

\begin{table*}[t]
\caption{\label{summary2}Continue of Table \ref{summary1}. The extrapolation value of AME20 is indicated with the \# symbol.}
\renewcommand{\arraystretch}{1.5}
\begin{ruledtabular}
\begin{tabular}{cllSScS} 
\textrm{Species}&
\multicolumn{1}{c}{\textrm{$\rho^2$}}&
\multicolumn{1}{c}{\textrm{$\delta (\rho^2) / \rho^2$}}&
\multicolumn{1}{c}{\textrm{ME~(keV)}}&
\multicolumn{1}{c}{\textrm{ME$_{\rm lit.}$~(keV)}}&
\multicolumn{1}{c}{\textrm{Ref.~of~lit.}}&
\multicolumn{1}{c}{\textrm{$\Delta {\rm ME}$~(keV)}} 
\\ \hline
\colrule
%
%
$A/q = 77.0$&&&&&\\
${}^{151}{\rm Nd}$${}^{40}{\rm Ar}_2^{3+}$&0.999801672(55)&$5.5 \times 10^{-8}$&-141025(42)&-141023.0(11)&AME20&-2(42)\\
${}^{151}{\rm Pr}$${}^{40}{\rm Ar}_2^{3+}$&0.999821116(45)&$4.5 \times 10^{-8}$&-136843(41)&-136860(12)&AME20&17(43)\\
${}^{1}{\rm H}{}^{12}{\rm C}{}^{32}{\rm S}_2^{1+}$&0.99983314(13)&$1.3 \times 10^{-7}$&-44752(16)&-44742.1032(26)&AME20&-10(16)\\
${}^{151}{\rm Ce}$${}^{40}{\rm Ar}_2^{3+}$&0.999846972(52)&$5.2 \times 10^{-8}$&-131282(41)&-131305(18)&AME20&23(45)\\
${}^{126}{\rm Sn}$${}^{14}{\rm N}_2^{2+}$&0.999897482(56)&$5.6 \times 10^{-8}$&-80279(28)&-80288(11)&AME20&9(30)\\

${}^{154}{\rm Sm}^{2+}$&0.999952088(25)&$2.5 \times 10^{-8}$&-72449(27)&-72455.6(13)&AME20&6(27)\\
${}^{154}{\rm Pm}^{2+}$&0.999981007(15)&$1.5 \times 10^{-8}$&-68303(27)&-68267(25)&AME20&-36(37)\\
${}^{154}{\rm Nd}^{2+}$&1&&&\multicolumn{1}{c}{\hspace{1.8em}$-65\hspace{0.2em}579.6(10 \rightarrow 130)$}&AME20&\\
${}^{154}{\rm Pr}^{2+}$&1.000052825(16)&$1.6 \times 10^{-8}$&-58005(27)&-58000.7(25)&CPT&-5(27)\\
${}^{153}{\rm Pr}$${}^{1}{\rm H}^{2+}$&1.000078794(75)&$7.5 \times 10^{-8}$&-54282(29)&-54258.5(24)&CPT&-23(29)\\
${}^{154}{\rm Ce}^{2+}$&1.000094033(58)&$5.8 \times 10^{-8}$&-52097(28)&-52068.9(24)&CPT&-28(28)\\
${}^{153}{\rm Ce}$${}^{1}{\rm H}^{2+}$&1.000126684(44)&$4.4 \times 10^{-8}$&-47415(27)&-47422.5(2.4)&CPT&7(27)\\
%
%
\hline
$A/q = 77.5$&&&&&\\
${}^{115}{\rm Cd}^{\rm m}$${}^{40}{\rm Ar}^{2+}$\footnote[1]{Admixture of the ground state and the isomeric state(s)}&0.999611874(70)&$7.0 \times 10^{-8}$&-122952(17)&-123124.38(65)&AME20&-9(17)\\
${}^{115}{\rm Pd}^{\rm m}$${}^{40}{\rm Ar}^{2+}$\footnotemark[1]&0.999664409(26)&$2.6 \times 10^{-8}$&-115370(15)&-115466(14)&AME20&6(20)\\
${}^{115}{\rm Rh}$${}^{40}{\rm Ar}^{2+}$&0.999706804(50)&$5.0 \times 10^{-8}$&-109252(16)&-109269.1(73)&AME20&17(18)\\

${}^{127}{\rm Sb}$${}^{14}{\rm N}_2^{2+}$&0.999902999(83)&$8.3 \times 10^{-8}$&-80939(19)&-80971.5(51)&AME20&33(19)\\
${}^{127}{\rm Sn}^{\rm m}$${}^{14}{\rm N}_2^{2+}$\footnotemark[1]&0.999925199(31)&$3.1 \times 10^{-8}$&-77735(15)&-77742.7(92)&AME20&3(18)\\

${}^{155}{\rm Sm}^{2+}$&0.999977444(16)&$1.6 \times 10^{-8}$&-70195(14)&-70191.2(13)&AME20&-4(14)\\
${}^{155}{\rm Pm}^{2+}$&1&&&\multicolumn{1}{c}{\hspace{1.5em}$-66\hspace{0.2em}940.0(47 \rightarrow 71)$}&AME20&\\
${}^{155}{\rm Nd}^{2+}$&1.000032340(15)&$1.5 \times 10^{-8}$&-62273(14)&-62283.8(9.2)&AME20&11(17)\\
${}^{155}{\rm Pr}^{2+}$&1.000079748(18)&$1.8 \times 10^{-8}$&-55431(14)&-55415(17)&AME20&-16(22)\\
${}^{154}{\rm Pr}$${}^{1}{\rm H}^{2+}$&1.000112383(71)&$7.1 \times 10^{-8}$&-50722(18)&-50711.7(2.5)&CPT&-10(18)\\
${}^{155}{\rm Ce}^{2+}$&1.00013418(19)&$1.9 \times 10^{-7}$&-47576(31)&-47780(300)\hspace{-1.5em}\#&AME20&204(302)\# \\
${}^{154}{\rm Ce}$${}^{1}{\rm H}^{2+}$&1.000153507(80)&$8.0 \times 10^{-8}$&-44787(18)&-44779.9(2.4)&CPT&-7(19)\\
\end{tabular}
\end{ruledtabular}
\renewcommand{\arraystretch}{1.0}
\end{table*}

\begin{table*}[t]
\setcounter{table}{2}
\caption{\label{w-average}Weighted average of the measured mass excess. The observed chi-square values of the weighted averages are indicated in column ``$\chi^2$." The literature mass excess values are shown in column ``ME$_{\rm lit}$" and their reference are indicated in the next column; AME20 \citep{Huang2021, Wang2021} and CPT \citep{Orford2022}. The differences between mass excess values, $\Delta {\rm ME} \equiv {\rm ME}_{\rm Weight. ~Ave.}-{\rm ME}_{\rm lit}$, are given in the last column. 
}
\renewcommand{\arraystretch}{1.5}
\begin{ruledtabular}
\begin{tabular}{cSlScS}
\multicolumn{1}{c}{\textrm{Species}}&
\multicolumn{1}{c}{\textrm{Weight. Ave.} (keV)}&
\multicolumn{1}{c}{\textrm{$\chi^2$}}&
\multicolumn{1}{c}{\textrm{ME$_{\rm lit}$} (keV)}&
\multicolumn{1}{c}{Reference of ME$_{\rm lit}$}&
\multicolumn{1}{c}{\textrm{$\Delta {\rm ME}$} (keV)} \\
\colrule
${}^{151}{\rm La}$ & -53542(17) & \hspace{-0.5em}0.07 & -53310(435) &AME20& -231(436) \\
${}^{151}{\rm Ce}$\footnote[1]{Admixture of the ground state and the isomeric state} & -61230(15) & \hspace{-0.5em}0.14 & -61225(18) &AME20& -5(23) \\
${}^{152}{\rm Ce}$ & -58891(17) & \hspace{-0.5em}0.00 & -58878.3(23) &CPT& -13(17) \\
${}^{153}{\rm Ce}$ & -54710(21) & \hspace{-0.5em}0.10 & -54711.5(24) &CPT& 2(21) \\
${}^{154}{\rm Ce}$ & -52082(15) & \hspace{-0.5em}0.40 & -52068.9(24) &CPT& -13(15) \\
${}^{153}{\rm Pr}$ & -61564(22) & \hspace{-0.5em}0.12 & -61547.5(24) &CPT& -17(22) \\
${}^{154}{\rm Pr}$ & -58009(15) & \hspace{-0.5em}0.03 & -58000.7(25) &CPT& -8(15) \\
${}^{151}{\rm Nd}$ & -70945(20) & \hspace{-0.5em}0.00 & -70943.2(11) &AME20& -2(20) \\
\end{tabular}
\end{ruledtabular}
\end{table*}

\begin{figure*}[t]
\setcounter{figure}{4}
\centering 
\begin{center}
\includegraphics[width=0.9\textwidth,  bb = 0 0 720 356, clip, trim=0 0 0 0]{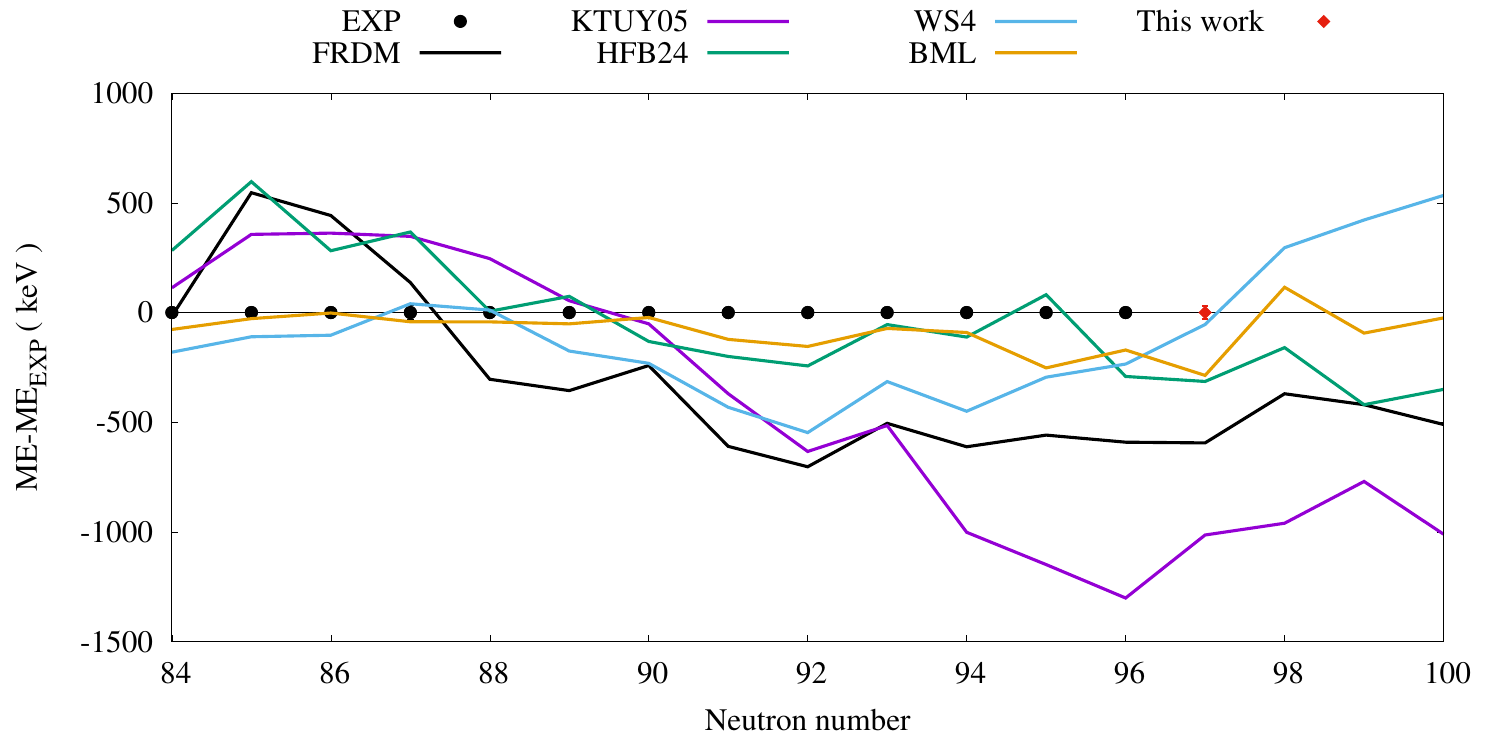}
\end{center}
\caption{Neutron-number dependence of  Ce-isotope mass excess. Each line indicates a theoretical prediction. See the text for their details. For $N \ge 98$, the 0 line corresponds to the extrapolation of AME20. }
\label{Ce-mass}
\end{figure*}

\begin{figure}[t]
\centering 
\begin{center}
\includegraphics[width=0.5\textwidth,  bb = 0 0 842 595, clip, trim=35 35 35 35]{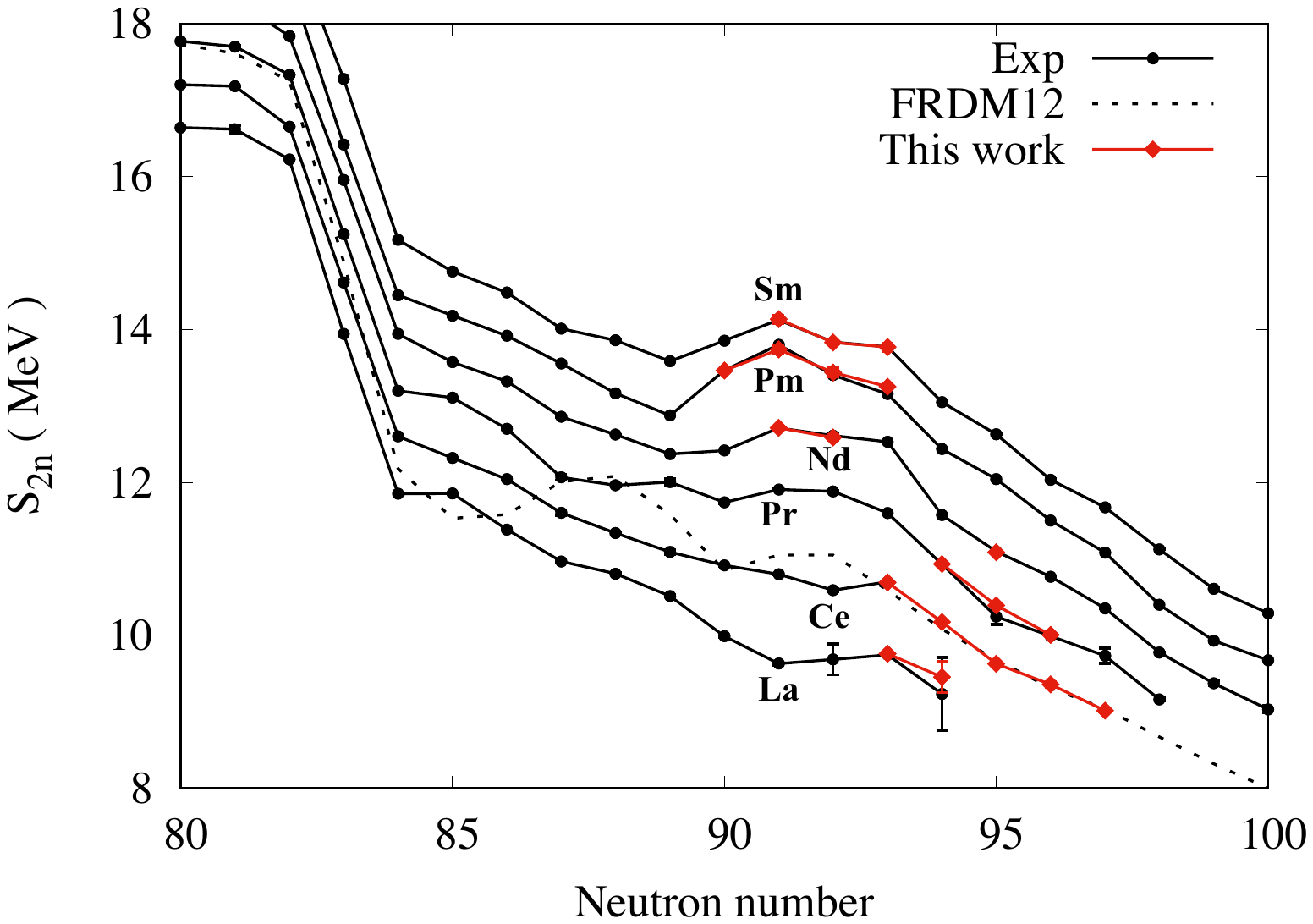}
\end{center}
\caption{Two neutron separation energy $S_{\rm 2n}$. The prediction of FRDM12, the dotted line, is given for Ce isotopes. }
\label{S2n}
\end{figure}

\begin{figure}[t]
\centering 
\begin{center}
\includegraphics[width=0.5\textwidth,  bb = 0 0 360 252, clip, trim=15 0 5 0]{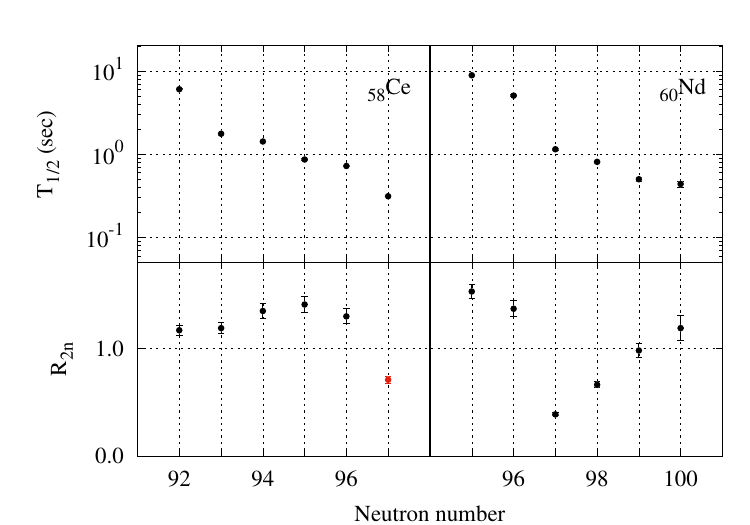}
\end{center}
\caption{Hafl-lives and $R_{\rm 2n}$-values (see Eq.~\ref{2nRatio}) of the Ce and Nd isotopes. The red point is the calculated value based on the ${}^{155}{\rm Ce}$ mass in the present measurement.}
\label{RQT}
\end{figure}

\subsection*{$\bm{A/q=75.5}$}
All observed peaks in the half-integer $A/q$ series belong to the $2+$ ions. We can reject the possibility of the $4+$ ions because no chemical element has a 4th ionization potential below the helium 1st ionization potential of 25 eV. In addition to atomic ions, the ions with adducts of $^{40}$Ar and $^{1}$H were identified. 

For ${}^{111}{\rm Pd}$, we determine a mass excess of ${\rm ME}({}^{111}{\rm Pd}) = -85866(30)~{\rm keV}$, which deviates from the AME value by $120(30)~{\rm keV}$, consistent with an admixture of the ground and the isomeric state of $E_{\rm X} = 172~{\rm keV}$. ${}^{151}{\rm Ce}$ also has a long-lived isomeric state, but its energy is still not determined. Any candidate of the isomeric state's peak cannot be found, and the peaks of ${}^{151}{\rm Ce}$ are treated as a single component in the fit process; the measured value includes the ambiguity that comes from this issue.

The uncertainty of the extracted mass excess value of $^{151}$La is $26~{\rm keV}$, while the literature value's uncertainty is $435~{\rm keV}$; we improved it by more than 15 times.

\subsection*{$\bm{A/q=76.0}$}
In this $A/q$ series, $^{40}{\rm Ar}_2$-attached $3+$ ions were observed in addition to the previous, $A/q=75.5$ case. The long-lived isomeric state of $E_{\rm X} = 38.5(26)~{\rm keV}$ exists in ${}^{112}$Rh \citep{Hukkanen2023a}. Two long-lived isomeric states of $E_{\rm X} = 150~{\rm keV}$ and unknown excitation energy have been reported for ${}^{152}{\rm Pm}$. The measured mass excess of ${}^{152}{\rm Ce}$ is consistent with our previous value, ${\rm ME}({}^{152}{\rm Ce}) = -58845(61)~{\rm keV}$ \citep{Kimura2019}. 

\subsection*{$\bm{A/q=76.5}$}
Three nuclides of ${}^{113}{\rm Ru}$, ${}^{125}{\rm Sn}$, and ${}^{125}{\rm In}$ have long-lived isomeric states in this $A/q$ series. Their excitation energies are $E_{\rm X}({}^{113}{\rm Ru}^{\rm m}) = 100.0(11)~{\rm keV}$ \citep{Hukkanen2023b}, $E_{\rm X}({}^{125}{\rm Sn}^{\rm m}) = 27.5~{\rm keV}$, and $E_{\rm X}({}^{125}{\rm In}^{\rm m}) = 360~{\rm keV}$, respectively. The peak of ${}^{125}{\rm In}^{\rm m}$ should be resolved from the ground state's one by considering the mass resolving power, but we cannot find it owing to the tail of the neighboring peak of ${}^{153}{\rm Sm}$. Thus, ${}^{113}{\rm Ru}$ and ${}^{125}{\rm Sn}$ are not taken into consideration for the modification of the mass reference error.

\subsection*{$\bm{A/q=77.0}$}
The ${}^{40}{\rm Ar}$-attached $2+$ ions were not observed in this $A/q$ series because there was an intense contaminant peak close to their predicted positions, and the IMD eliminated both. The isomeric state is reported for ${}^{154}{\rm Pm}$, but the excitation energy is unknown yet. The measured mass excess value shows lighter mass compared with the literature value evaluated based on the $\beta$-endpoint measurement of ${}^{154}{\rm Nd} (\beta^-) {}^{154}{\rm Pm}$ \citep{Greenwood1993}.

\subsection*{$\bm{A/q=77.5}$}
The long-lived isomeric states of $E_{\rm X} = 181~{\rm keV}$ and $E_{\rm X} = 41~{\rm keV}$ are known for ${}^{115}{\rm Cd}$ and ${}^{115}{\rm Pd}$, respectively. Therefore, they are not included in the mass correlation, but their measured mass values indicate that the contribution from the isomeric states is dominant in their peaks. ${}^{127}{\rm Sn}$ also has a long-lived isomeric state. However, its energy is as low as $5~{\rm keV}$, and ${}^{127}{\rm Sn}$ is not removed from the mass correlation for the reexamination of $\delta m_{\rm ref}( {}^{155}{\rm Pm}  )$. ${}^{127}{\rm Sb}$ is not included in the mass correlation due to the reason discussed below. 

The mass of ${}^{127}{\rm Sb}$ has been evaluated by using the mass of ${}^{127}{\rm Te}$ via the $\beta$-endpoint measurement \citep{Ragaini1967} and is determined to be ${\rm ME}({}^{127}{\rm Sb}) = -86665(19)~{\rm keV}$ by the direct measurement for the first time. We find the small peaks close to the predicted positions of ${}^{155}{\rm Ce}$. The fit results presented in Fig.~\ref{A=155++} show a similar intensity ratio to the others and also return similar values; the observed chi-square value of two measurements, $\rho^2_{602}({}^{155}{\rm Ce}) = 1.00013423(28)$ and $\rho^2_{604}({}^{155}{\rm Ce}) = 1.00013413(26)$, is $\chi^2({}^{155}{\rm Ce}) = 0.061$ and is sufficiently small to the 95\% confidence level of $\chi^2_{95\%} = 3.84$. Thus, we conclude these peaks are ${}^{155}{\rm Ce}$ and experimentally determine its mass for the first time; ${\rm ME}({}^{155}{\rm Ce}) = -47576(31)~{\rm keV}$.

For several isotopes, two chemical forms were measured independently.  The results are summarized in Table~\ref{w-average}. For ${}^{152-154}{\rm Pr}$, ${}^{1}{\rm H}$-attached $2+$ ions are observed, but ${}^{150,151}{\rm Pr}{}^{1}{\rm H}^{2+}$ ions are not and should exist in the TOF spectra of $A/q = 75.5$ and 76.5, respectively, by considering their expected yield at ${}^{252}{\rm Cf}$ spontaneous fission. The predicted positions of ${}^{150,151}{\rm Pr}{}^{1}{\rm H}^{2+}$ are close to ${}^{151,152}{\rm Ce}$'s ones and are overlapped with them. Based on the evaluated fission yields \citep{Katakura2012} and the intensity ratios of ${}^{152}{\rm Pr}{}^{1}{\rm H}^{2+}$/${}^{153}{\rm Pr}^{2+}$ and ${}^{153}{\rm Ce}$/${}^{153}{}\rm Pr$, the ratio of ${}^{150}{\rm Pr}{}^{1}{\rm H}^{2+}$/${}^{151}{\rm Ce}^{2+}$ (${}^{150}{\rm Pr}{}^{1}{\rm H}^{2+}$/${}^{151}{\rm Ce}^{2+}$) can be estimated to be $9.1 \times 10^{-3}$ ($2.4 \times 10^{-2}$) and would not affect the fit results. As shown in Table~\ref{w-average}, both results for atomic ions and molecules, which have no contaminants from ${}^{1}{\rm H}$-attached Pr, are consistent with each other. 

The measured mass excesses of ${}^{151-155}{\rm Ce}$ are plotted in Fig.~\ref{Ce-mass} as a function of neutron number with several theoretical predictions: the finite-range liquid-drop model (FRDM12) \citep{Moller2016}, the model of Koura-Tachibana-Uno-Yamada (KTUY05) \citep{Koura2005}, the Skyrme-Hartree-Fock-Bogoliubov mass formula with BSk24 parameter set (HFB24) \citep{Goriely2013}, the macroscopic-microscopic model with surface diffuseness correction (WS4) \citep{Wang2014}, and the Bayesian machine learning mass model (BML) \citep{Niu2022}. The variance of the predictions becomes more expansive in the region of $N > 90$, and the mass excesses of ${}^{152-155}{\rm Ce}$ locate the edge of this variance. No model can reproduce both values and trend of the ${}^{152-155}{\rm Ce}$ mass excesses, but FRDM12 and BML seem to be consistent with the result by considering only the trend. Two-neutron separation energies ($S_{\rm 2n}$) are plotted in Fig~\ref{S2n}, and the FRDM12 can reproduce the trend of Ce isotopes at $N \ge 94$. This means that the predicted neutron-capture reaction $Q$-values of several models agree with the new data, and the influence on the $r$-process calculations themselves is minor importance. 

The half-lives of neutron-rich lanthanoid isotopes were measured at RIKEN RIBF \citep{Wu2017}. The measured half-lives of the Ce and Nd isotopes are plotted in Fig.~\ref{RQT}, and the drops of the half-life trends at $N=97$ are observed for both elements. The authors conclude that the influence of the $\beta$-strength function would be dominant in the drop in the Nd case compared with the contribution of the $Q_{\beta}$-value: $T_{1/2} \propto Q_{\beta}^{-5}$. To see the evolution of the $\beta$-strength function, we introduce the ratio, $R_{\rm 2n}$, defined by 
\begin{equation}
R_{\rm 2n} (N) \equiv \frac{Q_{\beta}^{5}(N)T_{1/2}(N)}{Q_{\beta}^{5}(N-2)T_{1/2}(N-2)}.
\label{2nRatio}
\end{equation}
The obtained $R_{\rm 2n}$ of both isotope chains are also shown in Fig.~\ref{RQT}, and, for the Nd case, its trend changes at $N = 97$. The drops of the half-lives of Ce isotopes are observed at $N=93$ and $97$. The drop at $N=93$ would be explained by the $Q_{\beta}$-value systematic because there is no remarkable behavior in the $R_{\rm 2n}$ trend. In contrast, the new mass value of ${}^{155}{\rm Ce}_{97}$ leads the change of $R_{\rm 2n}$ trend at $N=97$, and we conclude the half-life drop at this point would be explained via the same way as the Nd case.

\section{Summary and Conclusions}

The masses of neutron-rich isotopes produced via spontaneous fission of ${}^{252}{\rm Cf}$ were measured using MRTOF-MS. For ${}^{155}{\rm Ce}$, the mass was experimentally determined for the first time with an error of $\delta m = 31~{\rm keV}$. A discrepancy between the experimental and literature values was found for the mass of ${}^{127}{\rm Sb}$, which was previously deduced through indirect measurements, and we propose new mass excess values for them. The mass excess value of $^{151}$La was determined with an error of $\delta m = 17~{\rm keV}$, which is approximately 25 times more precise than the literature value. When comparing with the theoretical predictions, there is no theoretical prediction that can explain both the values and trend of the ${}^{152-155}{\rm Ce}$'s mass excesses consistently. The MRTOF-MS's wide-range and simultaneous mass measurement facilitates check the consistency of the existing mass data. Without any correction, the part of the mass excess values extracted from the measured TOF ratios has inconsistency. The MRTOF-MS's high linearity suggests the possibility of a problem with either accuracy or precision or both evaluations in the existing mass data. This would mean the necessity of reexamining them. \\

\begin{acknowledgments}
This study was supported by the Japan Society for the Promotion of Science KAKENHI, Grant Number 17H06090 and 19K14750.
\end{acknowledgments}

\bibliography{n-richCe_mass}

\end{document}